\documentclass[superscriptaddress,showkeys,a4paper,twocolumn]{revtex4-1}
\usepackage[utf8]{inputenc}
\usepackage{amsmath}
\usepackage{amsfonts}
\usepackage{amssymb}
\usepackage{graphicx}
\usepackage[separate-uncertainty=true]{siunitx}
\DeclareSIUnit\fps{fps}
\DeclareSIUnit\pixel{pixel}

\usepackage{multirow}

\usepackage{tikz}
\usetikzlibrary{svg.path}
\usepackage{randtext}
\newcommand*{\svgat}{%
	\leavevmode
	\tikz[baseline=0pt, x=1pt, y=1pt, scale=1em/1000]\fill
	svg {
		M588 457v-241c0 -15 0 -66 35 -66c66 0 73 90 73 182c0 241 -171 351 -308
		351c-164 0 -307 -145 -307 -336c0 -179 130 -336 312 -336c94 0 187 24
		272 64c5 3 7 3 23 3h9c16 0 23 0 23 -10c0 -16 -120 -51 -146 -57c-64 -15
		-128 -22 -180 -22 c-200 0 -338 170 -338 358c0 199 150 358 333 358c161
		0 332 -133 332 -367c0 -107 -14 -210 -102 -210c-38 0 -90 19 -100 71c-30
		-42 -78 -71 -132 -71c-103 0 -198 93 -198 219s95 219 198 219c39 0 91
		-14 137 -77c6 -7 7 -8 23 -8h17c23 0 24 -1 24 -24zM519 262v170 c0 18 0
		21 -13 40c-36 56 -84 72 -116 72c-73 0 -132 -86 -132 -197s60 -197 132
		-197c20 0 71 6 115 69c14 21 14 25 14 43z
	}
	(current bounding box.west) ++(-56, 0) 
	(current bounding box.east) ++(56, 0) 
	;%
}
\newcommand*{\svgperiod}{%
	\leavevmode
	\tikz[baseline=0pt, x=1pt, y=1pt, scale=1em/1000]\fill
	svg {
		M192 53c0 -29 -24 -53 -53 -53s-53 24 -53 53s24 53 53 53s53 -24 53 -53z
	}
	(current bounding box.west) ++(-86, 0) 
	(current bounding box.east) ++(85, 0) 
	;%
}

\begin{document}
	
\title{Time-resolved quantification of plasma accumulation \\induced by multi-pulse laser ablation using self-mixing interferometry}

\author{S. Donadello}
\email[]{s\svgperiod\randomize{donadello}\svgat \randomize{inrim}\svgperiod it\\[1em]The Version of Record is available online at:\\ \href{https://doi.org/10.1088/1361-6463/abadc2}{doi.org/10.1088/1361-6463/abadc2}}
\affiliation{Department of Mechanical Engineering, Politecnico di Milano, Via La Masa 1, 20156 Milan, Italy}
\affiliation{Istituto Nazionale di Ricerca Metrologica, INRIM, Strada delle Cacce 91, 10135 Turin, Italy}
\author{V. Finazzi}
\affiliation{Department of Mechanical Engineering, Politecnico di Milano, Via La Masa 1, 20156 Milan, Italy}
\author{A. G. Demir}
\affiliation{Department of Mechanical Engineering, Politecnico di Milano, Via La Masa 1, 20156 Milan, Italy}
\author{B. Previtali}
\affiliation{Department of Mechanical Engineering, Politecnico di Milano, Via La Masa 1, 20156 Milan, Italy}

\begin{abstract}
	In this work a method based on self-mixing interferometry (SMI) is presented for probing the concentration of plasma plumes induced by multi-pulse laser ablation. An analytical model is developed to interpret the single-arm interferometric signal in terms of plasma electron number density. Its time dependence follows a power-law scaling which is determined by concurrent effects of plume accumulation and propagation. The model has been applied for the experimental study of plume formation at variable laser pulse frequencies on different materials. The plume expansion dynamics has been observed with high-speed imaging, and the SMI measurements allowed for a time-resolved estimation of the electron number density. The intrinsic advantages of the SMI technique in terms of robustness and low intrusiveness would allow for its usage as a fast diagnostic tool for the dynamical scaling of laser-induced plumes. Moreover it can be easily applied in laser-based manufacturing technologies where plasma concentration monitoring and control is important.
\end{abstract}


\maketitle

\section{Introduction}
The light-matter interaction drives a wide class of mechanisms which are of interest both for fundamental physics and technological applications. In such contexts, pulsed lasers represent powerful tools, since they allow to obtain high spatial and temporal energy densities capable of inducing material ablation \cite{chichkov_femtosecond_1996,shirk_review_1998}. These extreme conditions can lead to the formation of ablation plumes when highly energetic laser pulses hit a target material \cite{amoruso_characterization_1999,harilal_internal_2003}. The laser-induced plume is typically composed of a mixture of neutral vapors, solid particles, melt droplets, and plasma. In particular, free electrons, and ions can strongly interact with the laser beam photons and with the surrounding materials, decreasing the laser energy that would be absorbed by the target or introducing changes in the material surface properties \cite{fan_plasma_2002,bulgakova_impacts_2014}. Therefore, it would be useful to quantify the plume species around the working area in order to achieve a very precise control of certain laser-based processes. Such knowledge could be beneficial for example in pulsed laser deposition to control in a precise manner the amount and dynamics of particles reaching the target material surface, hence the process efficiency \cite{krishnaswamy_thinfilm_1989,hermann_plasma_1995,kwok_correlation_1997,thomas_pulse_2019}. Moreover, in laser processes such as laser microdrilling, laser texturing, and laser surface melting, the presence of plasma can alter the working surface, and it can perturb the processing optical beam by means of refractive defocusing or absorption caused by the plasma shielding effect \cite{mao_preferential_1996,marla_model_2014,cristoforetti_effect_2009,farrokhi_fundamental_2019,shaheen_experimental_2015,singh_effect_2017,stafe_theoretical_2007}. Accordingly the control of plume concentration can represent an important quality factor in precision machining.

Several techniques have been used to study and quantify the properties of the laser-induced plume. Among them, imaging methods allow for a direct observation of the plume evolution through the detection of plasma emission \cite{wen_laser_2007,farid_emission_2014,amoruso_features_2007,tao_effect_2006}, as well as shadowgraphy, fluorescence imaging or schlieren photography exploiting probe radiation across the ejected particles and vapors \cite{harilal_-_2017,porneala_time-resolved_2009,russo_time-resolved_1999,callies_time-resolved_1995,miyabe_ablation_2015}. Although high-speed cameras allow for good temporal and spatial resolutions, they essentially give only qualitative information about plume concentration. On the contrary, spectroscopic methods can be used to accurately quantify plasma species \cite{lednev_surface_2019,verhoff_dynamics_2012,corsi_effect_2005,claeyssens_plume_2002}; however the usage of spectrometers with high temporal resolution can be overkill and economically unfeasible for most of the industrial micromachining applications due to the instrument and data analysis costs. Instead, interferometric and holographic methods give rich information, allowing for time-resolved and quantitative measurements of the optical path difference introduced by the presence of species in atmosphere, and the consequent refractive index variations \cite{choudhury_time_2016,hough_enhanced_2012,sangines_two-color_2011,amer_shock_2008,pangovski_holographic_2016}. The drawback of common interferometric setups is their complexity and intrusiveness, which limits the integration in several manufacturing processes.

In comparison with other interferometric techniques, self-mixing interferometry (SMI), also known as feedback interferometry, represents a good candidate for the implementation of monitoring methods in industrial processes due to its robustness and small footprint, combined with a remarkably lower price \cite{taimre_laser_2015,donati_overview_2018,giuliani_laser_2002,li_laser_2017}. SMI allows to measure variations in the optical path which can be the outgrowth of changes in geometrical distance, refractive index, or both of them simultaneously. Traditionally SMI has been used to measure physical distances, velocities, or vibrations \cite{zabit_self-mixing_2013,magnani_spectral_2013,magnani_self-mixing_2012,donati_developing_2012}. In the field of laser machining, feedback interferometry has been demonstrated for probing the laser drilling depth \cite{mezzapesa_high-resolution_2011,mezzapesa_real_2012,demir_evaluation_2016}. Similarly, it is promising for sensing refractive index changes in laser-induced plumes \cite{colombo_self-mixing_2017,donadello_evolution_2018}. Such variations can be a consequence of changes in gas concentration, pressure, and temperature, with effects that are typically dominated by the electron density for ionized gases. Having a single-arm compact configuration, a feedback interferometer can be easily integrated on several laser-based applications, allowing for fast and non-invasive optical measurements. However to authors' knowledge a model for the interpretation of the self-mixing interferometric readout in terms of density for the plume species as a function of time was not present to date.

In the current work a method based on SMI is presented for the dynamical diagnosis of the laser ablation products within the plume volume. In particular, in the presence of plasma the measurement is dominated by the refractive index variation induced by the electron gas concentration. The approach which was introduced in our previous article on the topic has been generalized, solving some of its assumptions \cite{donadello_probing_2018}. In fact, although that preliminary paper demonstrated the possibility of using SMI to probe the ablation plume density, a quantitative and explicit interpretation for the nature of the plume species was not fully provided. Here an analytical model expresses the optical path difference in terms of electron density for the ionized gas accumulating within the expanding plume front during multi-pulse laser ablation. The corresponding power-law scaling parameters are experimentally accessible for a general description of the complex ablation plume dynamics as a function of time.

The model has been validated experimentally using a diode self-mixing interferometer, probing coaxially the ablation plume generated by nanosecond laser pulses with variable repetition rates on two different materials. Conditions of essentially superficial machining were considered in order to isolate the actual refraction index effect. A high-speed camera was used to synchronously observe the plume front evolution during the initial stages of plume formation, extracting the dynamical scaling parameters related to symmetry dimensionality and energy of the expansion wave. Such knowledge of the plume dynamics was used to calibrate the SMI model, allowing to estimate the plasma electron density from the interferometric optical path measurements with a high temporal resolution. A plasma accumulation effect was observed as the result of the multiple pulses. The measured power-law exponents were consistent with the model predictions, and the electron number density was of the order of \num{e24}--\SI{e25}{\per\meter\cubed} in agreement with literature for similar conditions. The proposed method represents an effective, low-cost, and robust possibility for the study of the laser-induced plume dynamics. Moreover, the results show that SMI can be used for a quantitative monitoring of the plasma electron density, in combination with a preliminary characterization of the model parameters. This might be functional to improve quality and efficiency in manufacturing processes such as laser micromachining and pulsed laser deposition.

\section{Model}
\begin{figure}
	\centering
	\includegraphics[width=0.85\linewidth]{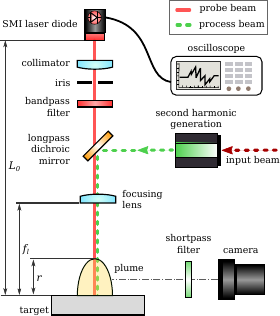}
	\caption{Sketch of the considered optical setup, including a high-power processing laser beam, a coaxial self-mixing interferometer, and a high-speed camera for the acquisition of the ablation-induced plume front.}
	\label{fig:setup}
\end{figure}

In this section a model is presented to link the optical path difference probed by a self-mixing interferometer and the refractive index variations in a laser ablation configuration as sketched in figure \ref{fig:setup}. Such perturbations are generated by the ablation-induced plume, and in particular can be related to the plasma electron density. The approach generalizes the model introduced in \cite{donadello_probing_2018}, overcoming some of the previous assumptions for the dynamical scaling parameters. These have been characterized experimentally by means of direct imaging of the expanding plume front, allowing for a quantitative interpretation of the SMI signal.

\subsection{Optical path difference}
After a high-power laser pulse hits a target surface at time $t=0$, the ablation process generates a plume which propagates from the ablation crater, forming jet-like structures through mechanisms of interest in many scientific and technological fields, and whose complex dynamics is object of several experimental and theoretical studies \cite{amer_comparison_2010,harilal_experimental_2012,farid_emission_2014,pangovski_holographic_2016,mayi_laser-induced_2019,ranjbar_plume_2020}. This ejection of droplets, nano-particles, neutral vapors, and ionized gases perturbs the optical path of any optical beam that crosses the ablation region, such as the processing laser itself or a dedicated interferometric probe. As a first approximation, the ablation-induced plume can be modeled as a volume with uniform refractive index $n(t)$, delimited by an expansion front whose position along the optical axis is expressed by the coordinate $r(t)$. Accordingly, the optical path length is calculated as
\begin{equation}
p(t) = n_0\,[L_0-r(t)] + n(t)\,r(t)\,,
\end{equation}
where $n_0$ is the refraction index of the unperturbed medium and $L_0$ is a reference geometrical length, assumed constant in time. For air $n_0-1 \simeq \num{2.8e-4}$, while $L_0$ can be taken as the distance between the probe source and the target surface as shown in figure \ref{fig:setup}.

It follows that, considering the case of a self-mixing interferometer coaxial to the plume expansion direction, the probed optical path difference can be expressed as
\begin{equation}
\delta p (t) = p(t)-p_0 = r(t)\, [n(t)-n_0]
\label{eq:optical-path-difference}
\end{equation}
relatively to the initially unperturbed optical path $p_0 = n_0L_0$. The interferometric measurement is therefore linked to the dynamical evolution of both the expansion front position $r(t)$ and the refractive index perturbation $n(t)-n_0$. The interferometric readout $\delta p (t)$ is typically expressed in units of interference fringe number, with each fringe corresponding to an optical path difference of half probe wavelength $\lambda_0/2$ \cite{donati_laser_1995}.

The refractive index of a dilute gas mixture can be expressed in first approximation as a linear combination of contributions given by different gas species and plasma constituents, i.e. electrons, ions and atoms \cite{ascoli-bartoli_wavelength_1960,anders_formulary_1990,ostrovskaya_holographic_2008}:
\begin{equation}
n(t) \simeq 1 + \sum_i K_i \rho_i(t)
\label{eq:refractive-mixture}
\end{equation}
where $K_i$ and $\rho_i(t)$ are the specific refractivity and number density for the $i$-th component, respectively.

For a neutral gas all the plume constituents are included in equation \eqref{eq:refractive-mixture}, which becomes the Gladstone--Dale formula for atoms. In the case of a plasma, for a probe wavelength in the visible spectrum and far from resonant electronic transitions, the free electron gas contribution is typically dominant and the effect of heavy ions and neutral atoms can be neglected \cite{walkup_studies_1986,breitling_shadowgraphic_1999,zhang_investigation_2009,salimi_meidanshahi_measurement_2013,amer_laser-ablation-induced_2009}. This is a realistic condition in pulsed laser ablation, where the generation of ionized gases is normally observed \cite{colombo_self-mixing_2017,donadello_probing_2018,zeng_experimental_2005,amoruso_features_2007}. Therefore, under the previous hypotheses, the refractive index of a dilute plasma depends only on the electron density $\rho_e(t)$ as
\begin{equation}
n(t) \simeq 1- \frac{\rho_e(t)}{2\rho_c}\,.
\label{eq:electron-refractive-index}
\end{equation}
The corresponding critical electron density $\rho_c$ is defined as
\begin{equation}
\rho_c = \frac{4\pi^2 \varepsilon_0 m_e c^2}{e^2 \lambda_0^2}\,,
\end{equation}
with $\varepsilon_0$ being the vacuum permittivity, $m_e$ the electron mass, $c$ the light speed, $e$ the elementary charge, and $\lambda_0$ the probe wavelength. In the current work $\lambda_0=\SI{785}{nm}$, hence $\rho_c=\SI{1.8e27}{\per\meter\cubed}$.

\subsection{Plume front expansion}
Several works have described the expansion of the laser ablation plume front in terms of the Sedov--Taylor theory \cite{gupta_direct_1991,callies_time-resolved_1995,bai_numerical_2016,campanella_shock_2019}, which assumes the formation of a blast wave as a consequence of an instantaneous explosion-like process. According to that theory the expansion of a blast wave formed at time $t=0$ can be described by means of a single coordinate $r(t)$, which scales with time as a power law:
\begin{equation}
r(t>0) = \epsilon t^{2\alpha}\,,
\label{eq:front-radius}
\end{equation}
where the scaling constant
\begin{equation}
\epsilon = \xi_0\left(\frac{E_0}{\rho_0}\right)^{\alpha}
\end{equation}
is determined by the shock wave energy $E_0$, the ambient gas density $\rho_0$, and the dimensionless constant $\xi_0 \sim 1$, which depends on the specific heat ratio. For air in standard atmosphere $\rho_0 \simeq \SI{1.2}{\kilogram\per\meter\cubed}$, while $E_0$ typically represents a small fraction of the laser pulse energy. The scaling exponent $\alpha$ depends on the blast wave expansion symmetry as
\begin{equation}
\alpha = \frac{1}{d+2}
\label{eq:front-exp}
\end{equation}
where $d$ represents the dimensionality:
\begin{equation}
\begin{split}
d=3 \quad &\rightarrow \quad \text{spherical wave}\\
d=2 \quad &\rightarrow \quad \text{cylindrical wave}\\
d=1 \quad &\rightarrow \quad \text{plane wave}.
\end{split}
\end{equation}

Different studies concerning laser ablation with strongly energetic pulses reported spherical wave expansions as expected for point-like explosions \cite{amer_shock_2008,porneala_time-resolved_2009,konig_plasma_2005,zeng_energy_2004}. Also plane wave or cylindrical expansions have been observed in the case of less abrupt ablations or longer laser pulses \cite{demir_investigation_2015,hough_enhanced_2012,jeong_propagation_1998}. Intermediate regimes or transients between different dimensionalities are also possible \cite{harilal_internal_2003,yavas_planar_1998,zeng_experimental_2005,corsi_effect_2005}, e.g. while observing the expansion on different timescales or during multi-pulse ablation. In our previous work a spherical wave with $d=3$ was assumed \cite{donadello_probing_2018}. However such hypothesis was based only on considerations from literature results, while it was not supported by direct experimental evidences. Here $d$ is not imposed to a fixed value in order to propose a more general and flexible model.

\subsection{Optical path for a single laser pulse}
As predicted by the blast wave theory and confirmed by several experiments, a low-pressure atmosphere is created behind the expanding shock wave front \cite{jeong_numerical_1998,tao_effect_2006,harilal_internal_2003,pangovski_control_2014}. It can be assumed that vapors and particles ejected from the material as a consequence of the ablation process get distributed within the rarefied plume volume $V(t)$ delimited by the high-pressure wave front at $r(t)$. Therefore it is reasonable to consider an effective refractive index $n(t)$ for the ablation plume, determined by the particle density averaged over the plume volume $V(t)$.

Considering the ablation-induced plasma, if the absolute number of free electrons produced by a single laser pulse is quantified by $N_e$, the average electron density $\rho_e(t)$ within ablation zone is
\begin{equation}
\rho_e(t>0) = \frac{N_e}{V(t)} \simeq \frac{N_e}{k_g r^d(t)\,l_T^{3-d}}\,.
\label{eq:electron-density}
\end{equation}
For this work it is assumed that the plume volume $V(t)$ scales with time as $r^d(t)$, thus depending on the expansion regime determined by $d$ and taking into account the system geometry. It follows that, from a dimensional point of view, $V(t)$ must be characterized by a characteristic length $l_T$ along the transverse direction relatively to the plume expansion, with $k_g$ being a geometrical constant. The transverse length $l_T$ is reasonably of the same order of magnitude of the laser spot interaction region, and here its time dependence is assumed negligible in first approximation. In the case of a plane wave, $k_g l_T^2$ is the transverse area of the wave front, with $k_g=\frac{\pi}{4}$; for a cylindrical expansion $l_T$ is the transverse height and $k_g=\frac{\pi}{2}$; in the case of a pure spherical wave such transverse dimension is not meaningful since it corresponds to a point-like explosion, while $k_g=\frac{2}{3}\pi$.

From the combination of equations \eqref{eq:electron-refractive-index} and \eqref{eq:electron-density} with equation \eqref{eq:optical-path-difference}, the optical path difference $\delta p(t)$ introduced by a single ablation pulse can be calculated at a generic time $t$ after the laser emission as
\begin{equation}
\delta p(t>0) \simeq - \frac{\rho_e(t)}{2\rho_c}r(t) = \zeta t^{\beta}\,,
\label{eq:optical-path-single}
\end{equation}
where the term $r(t)\,(1-n_0)$ has been neglected considering the refractive index of air $n_0 \simeq 1$. Accordingly, by taking the plume front expansion law of equation \eqref{eq:front-radius}, the optical path difference $\delta p(t)$ scales with time following a power law, whose scaling constant is
\begin{equation}
\zeta = -\frac{ N_e \epsilon^{1-d}}{2 k_g\rho_c l_T^{3-d}}
\label{eq:zeta-smi}
\end{equation}
with an exponent which depends only on the expansion symmetry dimensionality $d$ as
\begin{equation}
\beta = \frac{2(1-d)}{d+2}\,.
\end{equation}
Therefore, for an electron gas $\delta p(t>0)$ is negative, conversely to the case of a neutral gas. Moreover it must be noted that $\beta<0$ for $d>1$, which means that, excluding the case of a pure plane wave, the contribution of $\delta p(t>0)$ vanishes in time as it can be expected on long timescales for an expanding plume.

\subsection{Optical path for multiple laser pulses}
During a multi-pulse laser ablation a train of pulses is repeated at rate of $f_p=1/t_p$, with $t_p$ being the repetition period and with single-pulse duration $\tau_p \ll t_p$. As suggested by direct observations \cite{breitling_fundamental_2004,kraft_time-resolved_2020,pangovski_designer_2012}, in such conditions the plume dynamics resulting from ejection by subsequent laser pulses is chaotic, with an effective particle mixing and accumulation within the rarefied plume volume. The plume accumulation mechanism was recently confirmed also by numerical simulations, which highlighted that the multiple shock-waves generated by subsequent laser pulses merge after few pulses, forming a single plume enclosed in a low-pressure volume \cite{ranjbar_plume_2020}. Therefore, following the model proposed in our previous work \cite{donadello_probing_2018}, the total optical path difference $\Delta p(t)$ for a train of pulses can be seen as the superimposition of single-pulse contributions. In particular, considering long timescales $t \gg t_p$ and keeping the hypothesis of uniform refractive index for a dilute plume, a continuous-like accumulation process can be assumed. The single-pulse contribution estimated by equation \eqref{eq:optical-path-single} can be normalized to the pulse repetition period and referred to a relative timescale $t'$ spanning over the distributed ablation interval. It follows that $\Delta p(t)$ can be obtained by integrating $\delta p (t')/t_p$ from the ablation start in $0$ to the generic time $t$:
\begin{equation}
\Delta p(t>0) \simeq \int_{0}^{t}\frac{\delta p (t')}{t_p}dt' = \eta t^\gamma\,.
\label{eq:smi-delta-p}
\end{equation}

Therefore the optical path difference introduced by a multi-pulse ablation process scales with time as a power law, whose scaling constant $\eta$ depends on the repetition period $t_p$ and on the single-pulse scaling constant $\zeta$ as
\begin{equation}
\eta = \frac{\zeta}{t_p \gamma}\,.
\label{eq:eta-smi}
\end{equation}
The corresponding scaling exponent $\gamma$ can be expressed as
\begin{equation}
\gamma = \beta + 1 = 6\alpha - 1 = \frac{4-d}{d+2}\,.
\label{eq:exponents}
\end{equation}
Such relation is particularly interesting since it shows that the scaling exponents $\alpha$, $\beta$ and $\gamma$ are determined only by the expansion dimensionality $d$ introduced in equation \eqref{eq:front-radius}. In particular it links the dynamical scaling of the shock wave front $r(t)$ to a quantity which can be easily measured by means of interferometric methods, i.e. the optical path $\Delta p(t)$.

\begin{figure}
	\centering
	\includegraphics[width=1\linewidth]{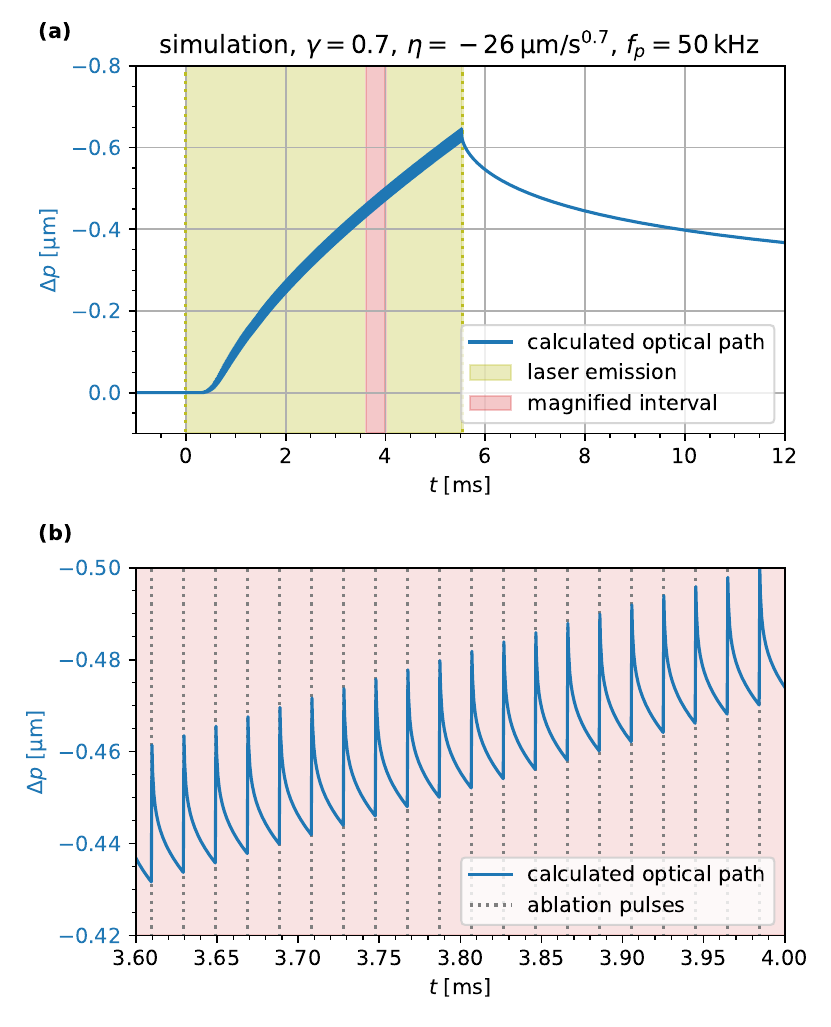}
	\caption{Numerical simulation of the optical path difference $\Delta p (t)$ during multi-pulse laser ablation, calculated as the cumulative sum of single-pulse contributions $\delta p(t)$ from the interaction model described in the text. The single-pulse contributions are magnified in (b). These were calculated at frequency $f_p=\SI{50}{\kilo\hertz}$ and normalized to the pulse peak power envelope of figure \ref{fig:laser-emission-prr}. For a qualitative comparison the values of $\gamma=0.7$ and $\eta=\SI{-26}{\micro\meter \per \second^{0.7}}$ correspond to the power law which fits the experimental data presented in figure~\ref{fig:ss-run1-17-plot-elab}.}
	\label{fig:ss-run1-17-plot-simulation}
\end{figure}

The behavior of $\Delta p (t)$ has been simulated numerically writing a Python code based on the Numpy library. The result is reported in figure \ref{fig:ss-run1-17-plot-simulation}. The values of $\eta$ and $\gamma$ taken as input parameters for the simulation were typical for the experimental conditions that will be discussed in the following sections. The calculation was performed by summing up single-pulse contributions $\delta p(t')$, with $t'$ delayed at pulse frequency $f_p = \SI{50}{\kilo\hertz}$ and considering $n_p=280$ discrete pulses. The single-pulse scaling constant $\zeta$ and exponent $\beta$ were derived from the values chosen for $\eta$ and $\gamma$, with a dimensionality $d=1.5$ calculated using equation \eqref{eq:exponents}. Moreover, $\zeta$ was normalized to a ramp from $0$ to $1$ shifted by about $20$ pulses: this was necessary to take into account of the initial transient in the laser pulse peak power, hence in the ablation efficiency, which was observed for the laser system used in the experiments. This allowed for a direct comparison with measurements. A \SI{4e-7}{\second} discretization was used, and divergent data corresponding to each pulse were omitted in the graph for clarity. From the magnification reported in figure \ref{fig:ss-run1-17-plot-simulation}(b) it can be seen that the sum of repetitive vanishing single-pulse contributions gives rise to an overall optical path difference which increases during the multi-pulse laser ablation. This reaches a maximum of the order of \SI{e-6}{\meter}, and decreases to zero on a longer timescale of few \si{ms} after the process end.

\subsection{Electron number density}
Considering the case of a dilute plasma, a time-explicit relation for the average electronic density $\rho_e(t)$ can be found by combining equations \eqref{eq:optical-path-difference}, \eqref{eq:electron-refractive-index}, \eqref{eq:front-radius} and~\eqref{eq:smi-delta-p}:
\begin{equation}
\rho_e(t>0) = -\frac{2 \rho_c}{\epsilon t^{2\alpha}} \, \Delta p (t) = - \frac{2 \rho_c \eta}{\epsilon}\, t^{\frac{2-d}{d+2}}  \,.
\label{eq:electronic-density}
\end{equation}
This shows that the interferometric signal strength $\eta$ is representative of the plasma electronic density. In particular, the temporal evolution of $\rho_e(t)$ can be extracted from the measurement of the optical path difference $\Delta p(t)$ during the ablation process, with $\Delta p(t>0)$ negative for an electron gas. Indeed, this does not require the knowledge of microscopic variables like $E_0$, $N_e$ and $l_T$, whose estimation is not always trivial.

\begin{figure}
	\centering
	\includegraphics[width=0.85\linewidth]{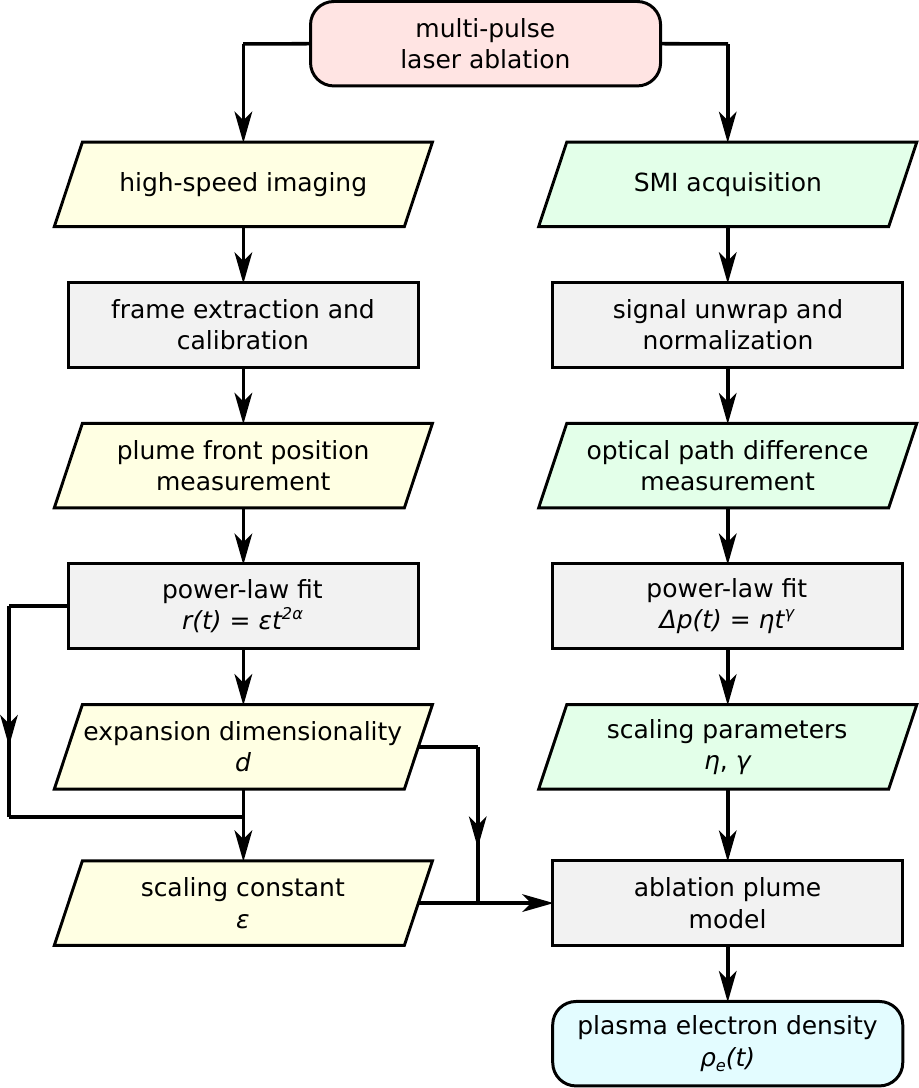}
	\caption{Steps followed for the calculation of plasma electron density from SMI measurements during laser ablation, modeled in terms of optical path difference introduced by the ablation plume. The dynamical parameters for the plume expansion are extracted by means of high-speed imaging.}
	\label{fig:flow-chart}
\end{figure}

The method used for the estimation of the electron number density is schematized in figure \ref{fig:flow-chart}. The model links an accessible optical measurement, such as $\Delta p(t)$ obtained via SMI, to $\rho_e(t)$. This approach might be applicable for a real-time control of the plasma plume concentration during a distributed process. For a quantitative measurement, a preliminary characterization of the plume front dynamics is required to calibrate the model for specific working conditions. In the current work $\alpha$ and $\epsilon$ are evaluated through power-law fit of the high-speed imaging measurements, while $\gamma$ and $\eta$ are similarly determined from the SMI acquisitions. 

\section{Experimental methods}
\subsection{Laser ablation}
The experimental setup used in the present work for the laser ablation process was the same described in \cite{colombo_self-mixing_2017}, based on a YLPG-5 pulsed fiber laser from IPG Photonics with \SI{532}{\nano\metre} second harmonic emission wavelength. The laser parameters are reported in table \ref{tab:process-par}. The laser source has a steady peak power of $P_\text{peak}=\SI{16}{\kilo\watt}$, and pulse duration is equal to $\tau_p= \SI{1.2}{\nano\second}$. The pulse repetition rate $f_p=t_p^{-1}$ was varied between \SI{50}{\kilo\hertz} and \SI{300}{\kilo\hertz}. The pulse train emission was triggered with a TTL signal having duration equal to $n_p t_p$, with $n_p=\num{280}$ the nominal ablation pulse number considered for all the tests. Experiments were carried out in standard atmosphere conditions, at about \SI{20}{\celsius} room temperature and relative humidity between \SI{30}{\percent} and \SI{60}{\percent}.

\begin{table}
	\centering
	\caption{\label{tab:process-par}Characteristics of the high-power pulsed laser used for the ablation experiments, with $f_p$ variable parameter.}
	\begin{tabular}{ll}
		\hline
		wavelength                  & \SI{532}{\nano\metre}                      \\
		peak power $P_\text{peak}$  & \SI{16}{\kilo\watt}                        \\
		pulse energy                & \SI{20}{\micro\joule}                      \\
		lens focal length $f_l$     & \SI{100}{\milli\metre}                     \\
		beam spot diameter          & \SI{53}{\micro\metre}                      \\
		pulse duration $\tau_p$     & \SI{1.2}{\nano\second}                     \\
		pulse number $n_p$          & \num{280}                                  \\
		pulse repetition rate $f_p$ & \num{50}, \num{150}, \SI{300}{\kilo\hertz} \\
		\hline
	\end{tabular}
\end{table}

\begin{figure}
	\centering
	\includegraphics[width=1\linewidth]{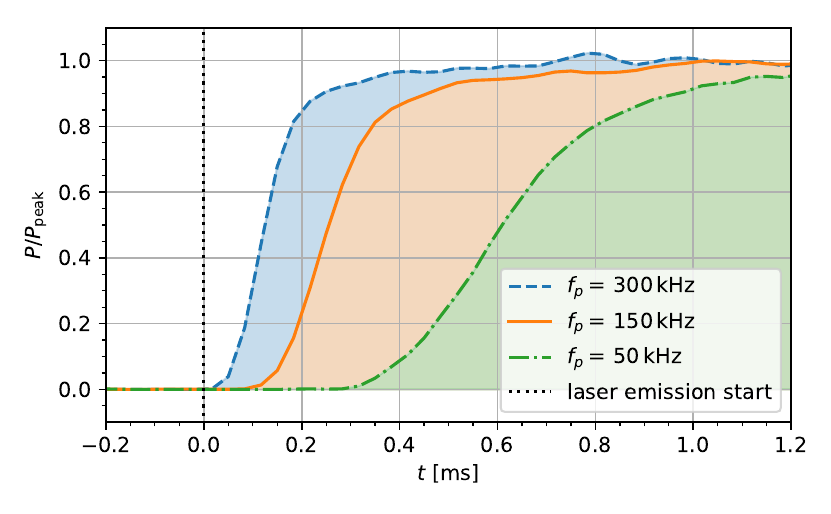}
	\caption{Characterization of the ablation laser pulse peak power $P$ as a function of emission time triggered in $t=0$, for the different values of pulse frequency $f_p$ and normalized to its steady value $P_\text{peak}=\SI{16}{\kilo\watt}$. Only the pulse train envelope is represented, since single pulses cannot be resolved in the graph within the whole interval.}
	\label{fig:laser-emission-prr}
\end{figure}

The actual pulse peak power $P$ of the processing laser was characterized using a silicon high-speed photodiode (Thorlabs FDS025), acquired as a function of emission time for the different pulse frequencies. The results for the envelope of $P$ are reported in figure \ref{fig:laser-emission-prr}, normalized to its steady value $P_\text{peak}$. This information was necessary to accurately interpret the trend of the optical path difference $\Delta p(t)$, since the emission of the high-power laser source is characterized by a delayed transient which cannot be neglected in the initial ablation interval. In particular, a delay corresponding to about $20$ pulses was observed in the laser emission.

The optical setup scheme was reported in figure \ref{fig:setup}. The optical path was characterized by a \ang{90} deflection of the process beam toward the target specimen with a long-pass dichroic mirror, having \SI{567}{\nano\metre} cut-off wavelength (Thorlabs DMLP567). An achromatic lens with focal length $f_l=\SI{100}{mm}$ (Thorlabs AC254-100-A-ML) was used to focus the processing beam on the target surface. The target was placed slightly out of focus in order to obtain a larger spot, \SI{1.4}{mm} farther, with a calculated diameter of \SI{53}{\micro\meter}. This ensured a lower laser intensity on the target, obtaining superficial machining conditions and less stringent requirements for the probe alignment.

Two different metallic materials were used as targets for the ablation plume formation: stainless steel (AISI 301) and commercially pure titanium (grade 2) sheets, with \SI{0.2}{mm} and \SI{0.3}{mm} thickness respectively. Ablation was replicated $5$ times for each combination of pulse frequency and target material. The chosen materials are representative of common applications, such as in laser micromachining or pulsed laser deposition. Moreover, they have similar physical properties, hence allow for a consistent result comparison. Nevertheless the analysis of results for different materials provides a convenient testbed to verify the plume model which has been presented. A further criterion for the material choice was represented by the plume front visibility with the direct high-speed imaging technique described in the following.

\subsection{Self-mixing interferometry}
A low-power GaAlAs laser diode (Hitachi HL7851G) was used for the self-mixing interferometer experiments, based on a multi-quantum well structure and with a built-in monitor photodiode. The SMI laser output was \SI{15}{\milli\watt} at $\lambda_0=\SI{785}{\nano\metre}$, and it was collimated using a \SI{10}{mm} focal length lens. The probe laser path was aligned coaxially with the optical axis of the process beam, and it was focused by the same achromatic lens. The SMI characteristics are reported in table \ref{tab:smi-pars}. The laser diode was placed at a reference distance $L_0$ from the target equal to about \SI{410}{\milli\metre} from the surface. A \SI{780}{nm} bandpass spectral filter (Thorlabs FBH780-10) was used to prevent spurious radiation to reach the laser diode.

\begin{table}
	\caption{\label{tab:smi-pars}Self-mixing interferometer characteristics.}
	\begin{tabular}{ll}
		\hline
		wavelength $\lambda_0$             & \SI{785}{\nano\metre}        \\
		power                              & \SI{15}{\milli\watt}         \\
		lens focal length $f_l$            & \SI{100}{\milli\metre}       \\
		\multirow{2}*{beam spot diameters} & \SI{42}{\micro\metre} (fast axis) \\
                                  & \SI{26}{\micro\metre} (slow axis) \\
        \hline
	\end{tabular}
\end{table}

When a train of high-power laser pulses hits the target, a plume is generated as a consequence of the ablation process and starts to propagate. The coaxial SMI probe beam interacts with such expanding ablation plume before reaching the target surface. Then, the portion of radiation scattered or reflected by the surface travels back through the same optical path and enters in the laser diode, where self-mixing interference takes place \cite{giuliani_laser_2002,donati_developing_2012,taimre_laser_2015}. An iris was used in the experiments to limit the optical feedback and operate the self-mixing interferometer in the moderate regime. In such condition the sawtooth-like signal modulation allows to distinguish the sign of the corresponding optical path difference from the slope of the voltage jump associated to each interference fringe.

The voltage signal $v_\text{SMI}(t)$ coming from the integrated monitoring photodiode was acquired with an oscilloscope having \SI{350}{\mega\hertz} bandwidth and \SI{50}{Msps} sampling rate (Rigol MSO4024). The signal analysis followed the methodology described in \cite{donadello_probing_2018}. The signal noise was reduced by a low-pass filter with \SI{500}{\kilo\hertz} cutoff. Then, an automatic algorithm was used to remove the signal offset and unwrap the interference fringes, whose amplitudes were normalized to half probe wavelength $\lambda_0/2$ obtaining a continuous optical path signal $\Delta p(t)$. In the considered setup a reduction in the optical path corresponded to a positive voltage difference, thus the unwrapped $v_\text{SMI}(t)$ and $\Delta p(t)$ signals had opposite signs.

\subsection{High-speed imaging}
The plume spatial evolution was observed with a high-speed CMOS camera placed near the target, transversely to the ablation setup optical axis. The camera was a Fastcam Mini AX200 from Photron, whose parameters are reported in table \ref{tab:camera-pars}. The resolution was $128\times \SI{208}{\pixel}$, with a image pixel magnification corresponding to \SI{2.36}{\micro\metre}. The acquisition frame rate was \SI{e5}{\fps}, and it was triggered with the processing laser control signal. A shortpass optical filter with \SI{500}{nm} cutoff (Edmund Optics 47-287) prevented the high-power processing laser radiation to reach the camera sensor.

\begin{table}
	\caption{\label{tab:camera-pars}High-speed imaging parameters.}
	\begin{tabular}{ll}
		\hline
		shutter time      & \SI{10}{\micro\second}            \\
		frame rate        & \SI{e5}{\fps}                     \\
		resolution        & $\num{128}\times\SI{208}{\pixel}$       \\
		calibration ratio & \SI{2.36}{\micro\metre\per\pixel} \\ 
		\hline
	\end{tabular}
\end{table}

The plume front detection was performed offline for the frame series acquired with the high-speed camera. Unlike techniques which use an additional probe beam, such as shadowgraphy or resonant imaging, in the current work the camera was used for a direct observation of the plume radiation. Indeed, the plume front expansion can be identified by the micro- and nano-sized particles ejected from the target during the initial ablation stages by melting or vaporization processes \cite{yoo_evidence_2000,pangovski_control_2014}. In particular, the black-body thermal emission of such bigger particles could be detected by the camera \cite{claeyssens_plume_2002,amoruso_features_2007}, with the particle heating being favored by the ongoing interaction with the ablation laser pulse train. The acquisitions suggested that these particles propagate within the plume volume, accumulating behind the high pressure shock wave formed after the process ignition. Accordingly, for each test the plume extension $r(t)$ was measured at different instants as the distance between the ablation crater on the target surface and the farther particle front. The imaging technique presented here is quite simple, but it was not applicable to other materials which were previously taken into account \cite{donadello_probing_2018}, such as TiN ceramic coating or copper: their image acquisitions showed a weaker signal for the ejected particles, and a clear plume front was not distinguishable, excluding the possibility of considering them for a quantitative study. Probably this was consequence of the different physical properties and ejection mechanisms associated to those materials, which caused a weaker thermal radiation intensity scattered from the ejected particles. Such limit might be overcome by other imaging methods, e.g. by observing the transmission image using external illumination.

\section{Experimental results}
\subsection{Ablation plume formation}
\begin{figure*}
	\centering
	\includegraphics[width=0.7\linewidth]{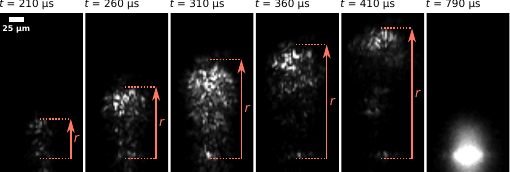}
	\caption{Measurement of the expanding plume front coordinate $r(t)$ for different frames acquired with high-speed imaging during the ablation of a stainless steel specimen at pulse frequency $f_p=\SI{150}{\kilo\hertz}$. Time is relative to the nominal laser trigger in $t=0$, although the actual ablation process was delayed due to the emission transient shown in figure \ref{fig:laser-emission-prr}. The last frame presents the plasma emission appearing on a longer timescale.}
	\label{fig:frames}
\end{figure*}

An example of high-speed acquisition for the ablation plume is reported in figure \ref{fig:frames}. After a time delay characteristic of the laser source emission, of the order of \SI{0.2}{ms}, a preliminary cloud of particles was ejected from the target surface. As a consequence of the subsequent laser pulses other particles were ejected and accumulated behind a common expanding front. The front dynamics showed a mainly axial propagation, while the transverse expansion appeared limited on the considered timescale, with a half-dispersion angle which has been estimated as of the order of \ang{10}. After the initial instants the plume propagation underwent deceleration, and $r(t)$ was measured up to about \SI{0.3}{mm} before it exited the camera field of view. Accordingly, the following analysis assumed the absence of significant alterations in the propagation mechanism on longer time intervals.

Besides the plume front propagation detected by means of the thermal radiation scattered by the bigger plume particles, a different kind of emission was typically observed on a longer timescale of the order of \SI{1}{ms}, as visible in the last frame of figure \ref{fig:frames}. Its continuous structure could be attributed to the ionized gas emission, which becomes visible when the plasma density reaches the camera sensitivity threshold in the region around the target surface, where locally higher temperature and density can be expected.

\begin{figure}
	\centering
	\includegraphics[width=0.85\linewidth]{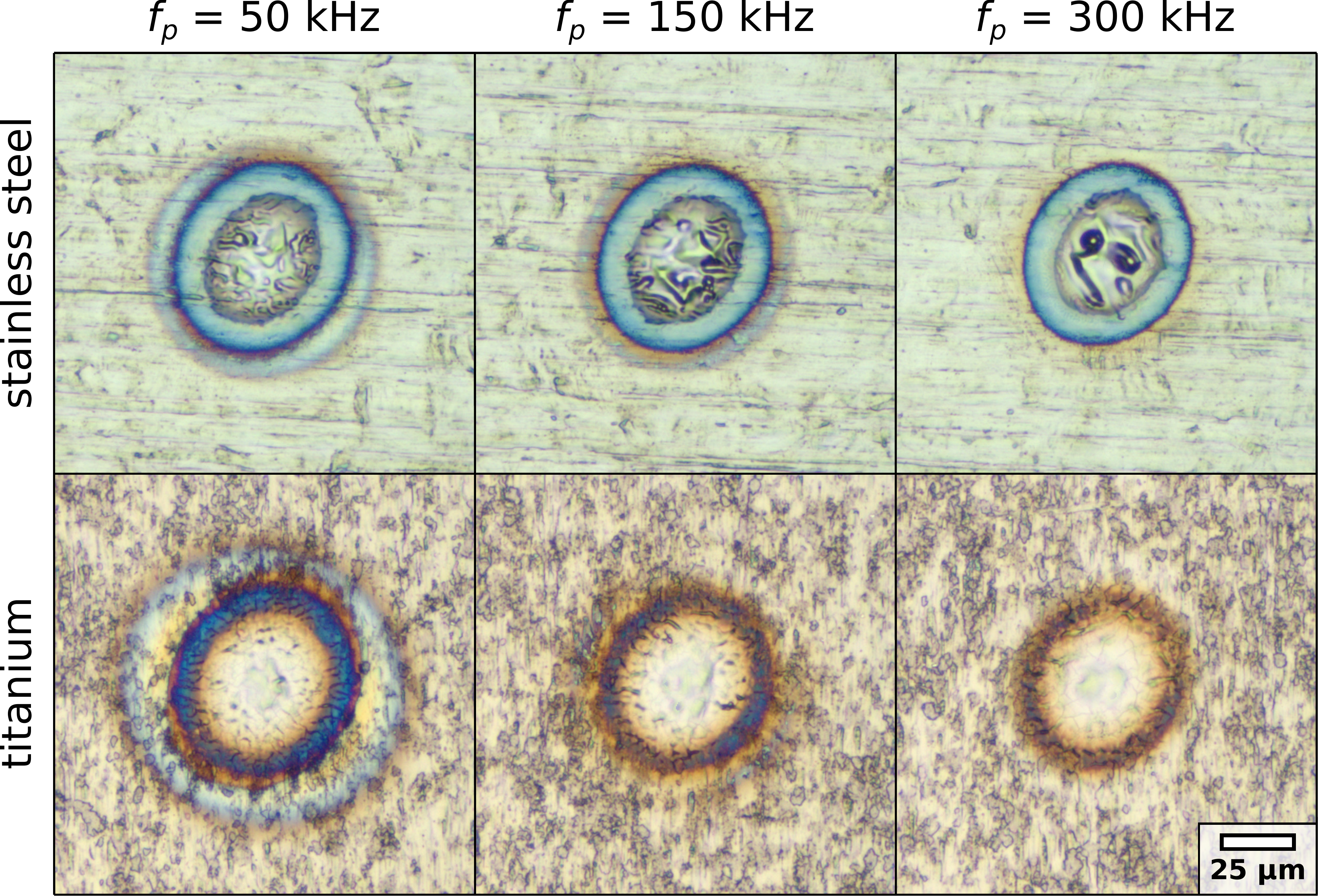}
	\caption{Example of ablation specimens obtained in the different experimental conditions.}
	\label{fig:holes}
\end{figure}

Figure \ref{fig:holes} reports optical microscopy images of ablation targets for each processing condition. Different oxidation effects can be observed on the target surface. Slower pulse repetition rates corresponded to longer interactions with laser radiation and plasma, hence to larger affected zones. Almost superficial machining was obtained for the considered process parameters and materials. Therefore, target deformation could be neglected, and the SMI probe interaction can be mainly ascribed to the refraction index variations induced by the ablation plume.

\subsection{Plume front and optical path difference}
\begin{figure}
	\centering
	\includegraphics[width=1\linewidth]{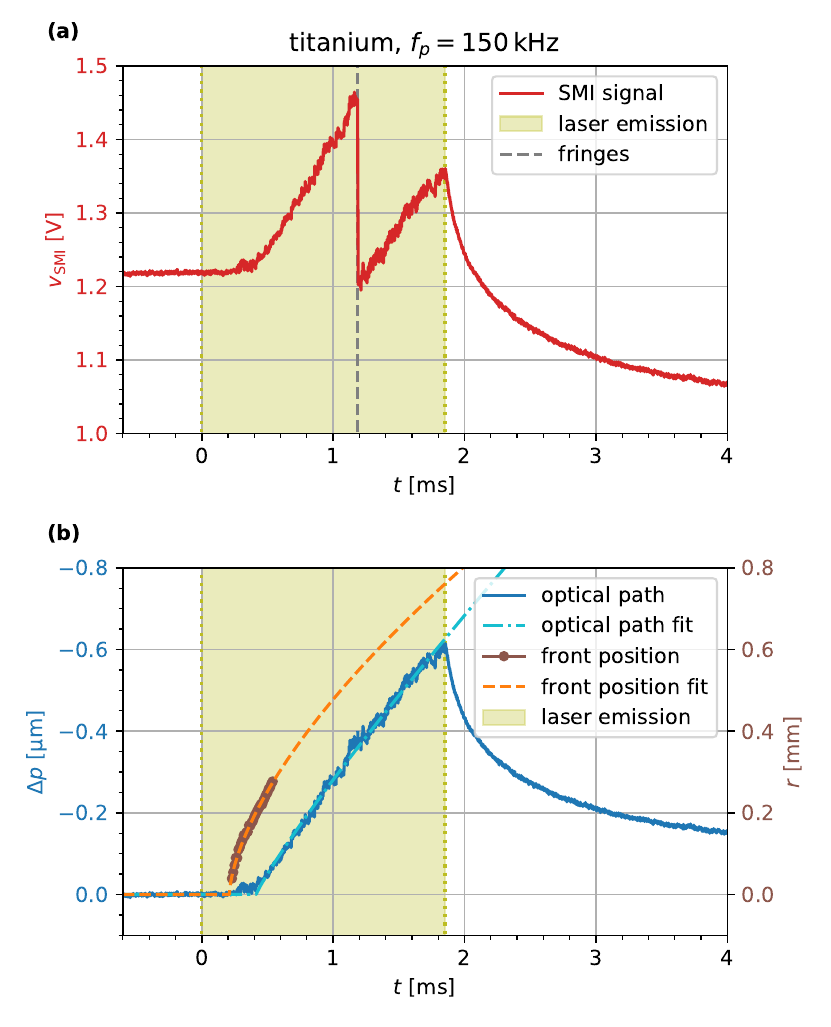}
	\caption{(a) SMI signal acquired during a titanium sample ablation at \SI{150}{\kilo\hertz} laser pulse frequency. Fringe identification is reported, and time is referred to the nominal ablation start in $t=0$. (b) Corresponding unwrapped optical path difference (left scale) and plume front position (right scale), with the respective fitting power-law curves.}
	\label{fig:ti-run3-23-plot-elab}
\end{figure}

An example of experimental data series is reported in figure \ref{fig:ti-run3-23-plot-elab} for a titanium sample with ablation pulse repetition rate equal to \SI{150}{\kilo\hertz}. The raw SMI voltage $v_\text{SMI}(t)$ is reported in figure \ref{fig:ti-run3-23-plot-elab}(a). The analysis algorithm identified the interferometric fringes and unwrapped the voltage fringe jumps, normalizing their amplitude to $\lambda_0/2$ to obtain the optical path difference $\Delta p(t)$.

The evolution of $\Delta p(t)$ and $r(t)$ with time can be compared on the same timescale of figure \ref{fig:ti-run3-23-plot-elab}(b) for the considered example. In general, after a delay of the order of \num{0.1}--\SI{0.5}{ms} relatively to the trigger signal for the ablation laser emission, the formation of a plume front was observed with a subsequent propagation from the target surface. Simultaneously, a negative optical path difference was detected, whose absolute magnitude increased until the multi-pulse ablation was interrupted, typically reaching a maximum between \SI{0.1}{\micro\meter} and \SI{1}{\micro\meter}. Then, after the laser emission finished, $\Delta p(t)$ decreased returning close to $0$ in few \si{ms}. This is in accordance with the model proposed for the optical path difference in the case of a ionized ablation plume, as predicted by equation \eqref{eq:smi-delta-p}.

The $r(t)$ and $\Delta p(t)$ curves within the laser emission interval were fitted to a power law to measure the dynamical scaling parameters of equations \eqref{eq:front-radius} and \eqref{eq:smi-delta-p}, $\epsilon$ and $\alpha$, $\eta$ and $\gamma$, respectively. A time offset parameter was included to take into account the variable delay in the laser source emission relatively to the trigger in $t=0$, whose characterization was reported in figure \ref{fig:laser-emission-prr}. The effect of the laser emission transient was clearly visible as a delayed $r(t)$ curve, and as a $\Delta p(t)$ smoothing in the initial ablation stage.

\begin{figure}
	\centering
	\includegraphics[width=1\linewidth]{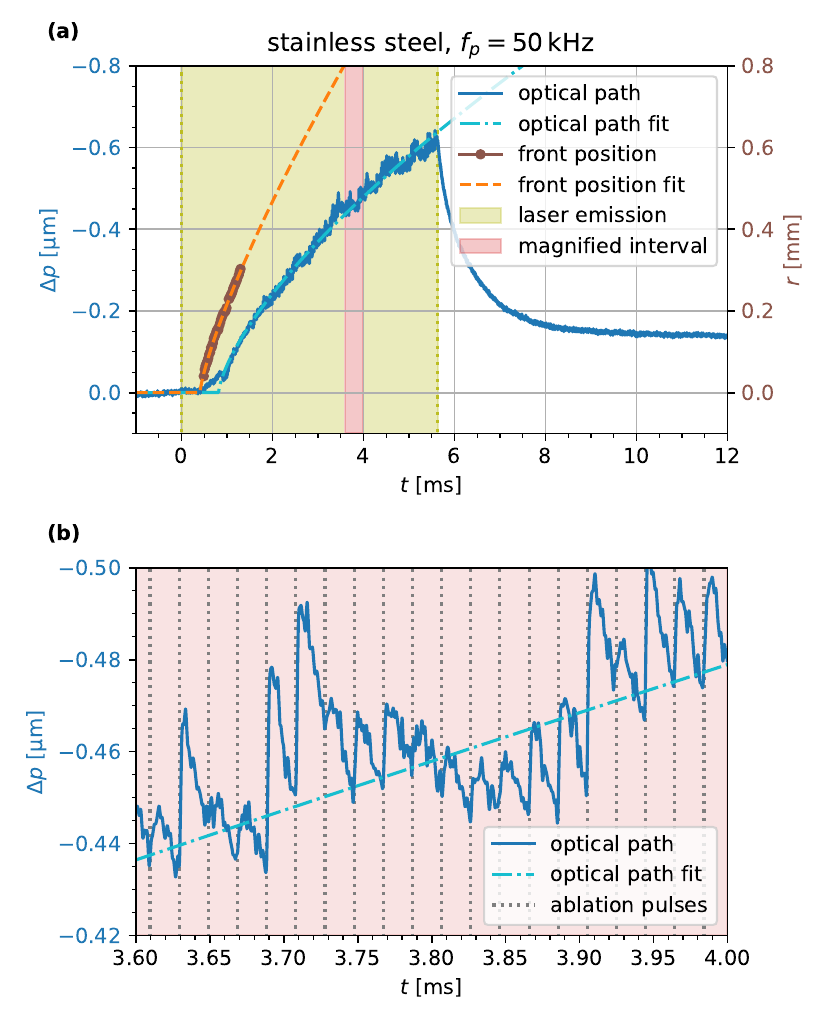}
	\caption{(a) Optical path difference and plume front position as a function of ablation time for a stainless steel target. The interval highlighted in red is magnified in (b), where the vertical lines represent the ablation pulses at $\SI{50}{\kilo\hertz}$ rate, to be compared with the calculation reported in figure \ref{fig:ss-run1-17-plot-simulation}.}
	\label{fig:ss-run1-17-plot-elab}
\end{figure}

Figure \ref{fig:ss-run1-17-plot-elab} reports another example of data series acquired for the ablation of a stainless steel specimen with $f_p=\SI{50}{\kilo\hertz}$. In particular, figure \ref{fig:ss-run1-17-plot-elab}(b) represents the magnification of an example interval of $\Delta p(t)$, highlighting the presence of small amplitude impulses whose periodicity matches the pulse repetition rate $f_p$. It can be seen that there is a good qualitative agreement with the simulated behavior reported in figure \ref{fig:ss-run1-17-plot-simulation}, which was calculated using the scaling parameters $\eta$ and $\gamma$ fitting the experimental $\Delta p(t)$ curve. This confirms the model interpretation in terms of plume accumulation. In fact, each ablation pulse generates a small contribution $\delta p(t)$ which tends to vanish due to its dilution within the expanding plume volume, but which gives an overall increasing trend for $\Delta p(t)$ since the process is sustained by a series of multiple laser pulses. Finally, these observations demonstrated that SMI is in principle applicable also for the dynamical study of plumes induced by single laser pulses; however high-speed and low-noise electronics should be developed to acquire small photodiode signal variations on fast \si{ns}-timescales, which cannot be easily distinguished with the current setup on a limited dynamical range.

\subsection{Dynamical scaling parameters}
\begin{figure}
	\centering
	\includegraphics[width=1\linewidth]{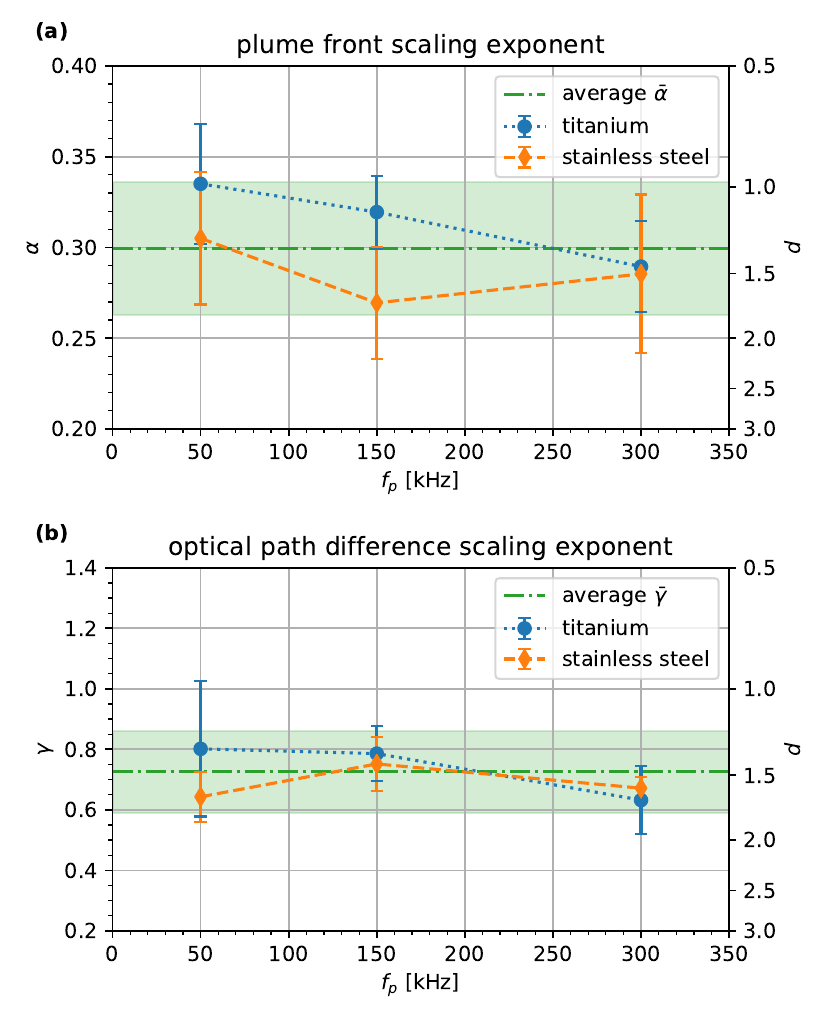}
	\caption{Power-law exponents fitting the experimental curves for the plume front expansion (a) and the SMI optical path difference (b), averaged and grouped by experimental conditions. Error bars represent standard deviations. The average over the whole data set is also reported, with the shadowed interval representing its standard deviation. The right scale represents the corresponding expansion dimensionality $d$ calculated from the model.}
	\label{fig:experimental-exp-fit}
\end{figure}

An initial analysis was performed on the dynamical scaling exponents $\alpha$ and $\gamma$, measured by fitting power laws to the $r(t)$ and $\Delta p(t)$ experimental curves, respectively. The results are reported in figure \ref{fig:experimental-exp-fit}, averaging the $5$ replicates taken for each process condition and excluding experimental failures. A clear trend of $\alpha$ and $\gamma$ with the pulse frequency was not observed in the considered interval of $f_p$. Moreover, all the exponent data points overlap each other within their uncertainties, hence differences between the considered materials are not evident from the measurements.

\begin{figure}
	\centering
	\includegraphics[width=1\linewidth]{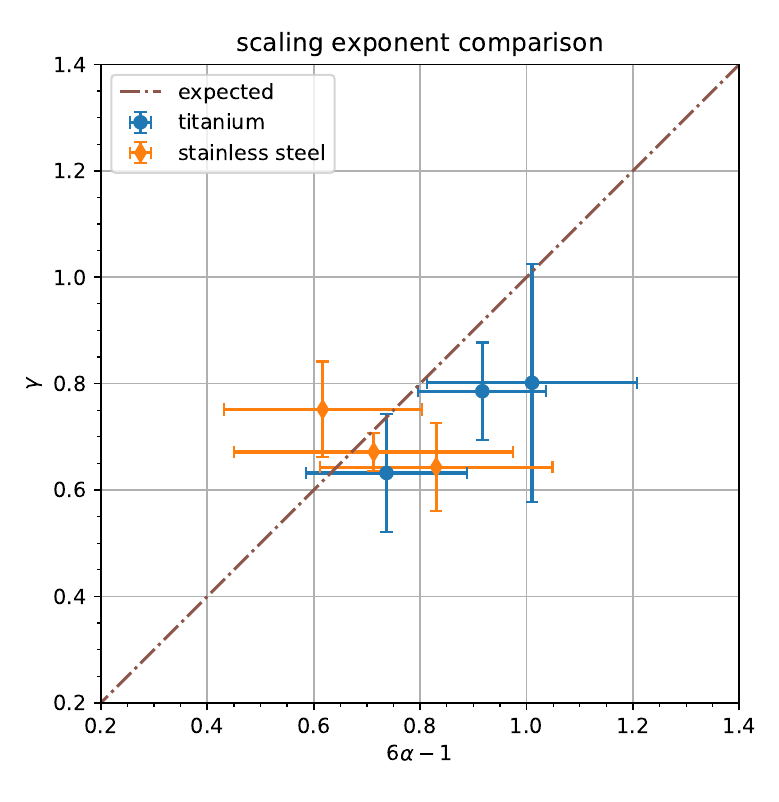}
	\caption{Comparison between the measured SMI scaling exponents $\gamma$, averaged for each experimental condition, and the corresponding values of $6\alpha-1$ calculated from the measured plume front expansion exponents. Diagonal line represents the correspondence expected from equation \eqref{eq:exponents}.}
	\label{fig:exponents-scatter}
\end{figure}

From a different point of view, equation \eqref{eq:exponents} links the plume front exponent $\alpha$ with the SMI optical path exponent $\gamma$. Therefore, it is possible to compare the experimental value of $\gamma$ with the corresponding value derived from the model as $6\alpha-1$. The results are reported in figure \ref{fig:exponents-scatter}. It can be seen that the experimental points overlap within the respective error bars with the diagonal line predicted by the model, thus the observed discrepancies can be ascribable to statistical dispersion relative to a common value determined only by the expansion mechanism.

The previous observations suggest that $\alpha$ and $\gamma$ do not depend on target material and pulse repetition rate, as assumed in the model. Accordingly, the scaling exponents of all experiments can be reasonably averaged, finding the values reported in table \ref{tab:smi-exponents}. The average dimensionality corresponding to the plume front exponent can calculated using equation \eqref{eq:exponents} as
\begin{equation}
\bar{d}(\alpha)=\frac{1}{\bar{\alpha}}-2=\num{1.4\pm0.4}\,,
\end{equation}
which has a good correspondence with the dimensionality calculated from the optical path exponent as
\begin{equation}
\bar{d}(\gamma)=\frac{4-2\bar{\gamma}}{\bar{\gamma}+1}=\num{1.5\pm0.3}\,.
\end{equation}
The agreement between the two dimensionality measurements supports the validity of the model hypothesis. Moreover the results indicate that the effective plume dynamics is determined by an intermediate behavior between the planar and cylindrical expansion regimes.

\begin{table}
	\caption{\label{tab:smi-exponents}Measured scaling exponents $\alpha$ and $\gamma$, averaged for the different experimental conditions. Last column reports the corresponding SMI exponent calculated from the front wave exponent according to the model. Last row reports the average values for the whole data set.}
	\begin{tabular*}{\columnwidth}{lc@{\hskip 1.6ex}c@{\hskip 1.6ex}c@{\hskip 1.6ex}c}
		\hline
		& $f_p$                 & $\alpha$         & $\gamma$         & $6\alpha-1$       \\ \hline
		\multirow{3}*{titanium}        & \hspace{1ex}\SI{50}{\kilo\hertz}  & \num{0.34+-0.03} & \num{0.80+-0.22} & \num{1.01+-0.2}   \\
		& \SI{150}{\kilo\hertz} & \num{0.32+-0.02} & \num{0.79+-0.09} & \num{0.92+-0.12}  \\
		& \SI{300}{\kilo\hertz} & \num{0.29+-0.03} & \num{0.63+-0.11} & \num{0.74+-0.15}  \\ \hline
		\multirow{3}*{stainless steel} & \hspace{1ex}\SI{50}{\kilo\hertz}  & \num{0.31+-0.04} & \num{0.64+-0.08} & \num{0.83+- 0.22} \\
		& \SI{150}{\kilo\hertz} & \num{0.27+-0.03} & \num{0.75+-0.09} & \num{0.62+-0.19}  \\
		& \SI{300}{\kilo\hertz} & \num{0.29+-0.04} & \num{0.67+-0.04} & \num{0.71+-0.26}  \\ \hline
		average                        & --                    & \num{0.30+-0.04} & \num{0.73+-0.14} & \num{0.80+-0.21}  \\
		\hline
	\end{tabular*}
\end{table}

\begin{figure}
	\centering
	\includegraphics[width=1\linewidth]{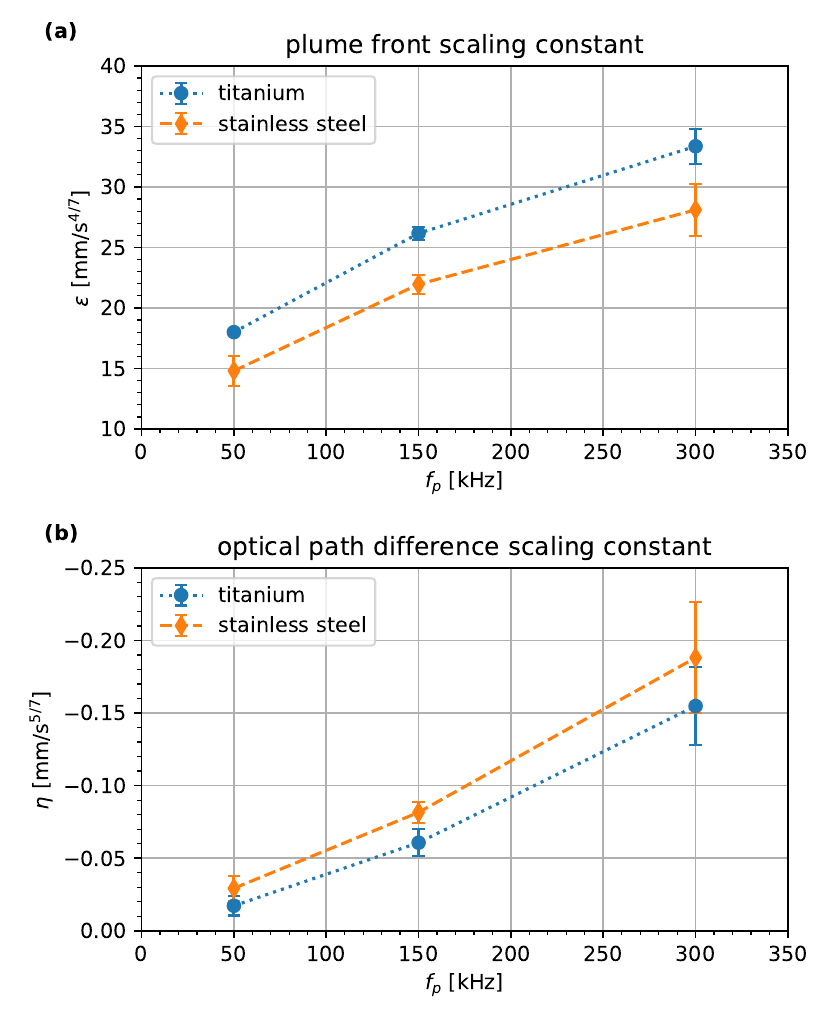}
	\caption{Power-law scaling constants fitting the experimental curves for the plume front expansion (a) and the SMI optical path difference (b), averaged and grouped by experimental conditions. Dimensionality was fixed to $d=1.5$.}
	\label{fig:experimental-amp-fit}
\end{figure}

Due to the variability observed for the scaling exponents, that strongly influences the power-law fit output, the procedure has been repeated by fixing the dimensionality to $d=1.5$ in agreement with the average experimental value $\bar{d}$. This allowed for a quantitatively consistent comparison between the different experimental conditions for the $r(t)$ and $\Delta p(t)$ scaling constants, $\epsilon$ and $\eta$, respectively. The results are reported in figure \ref{fig:experimental-amp-fit}. The general trend for the plume front expansion shows that the scaling constant $\epsilon$ grows with $f_p$, with a faster propagation observed for titanium. Also the optical path difference scaling constant $\eta$ increases with the pulse repetition rate, with a faster increase for stainless steel.

\subsection{Electron number density}
\begin{figure*}
	\centering
	\includegraphics[width=0.9\linewidth]{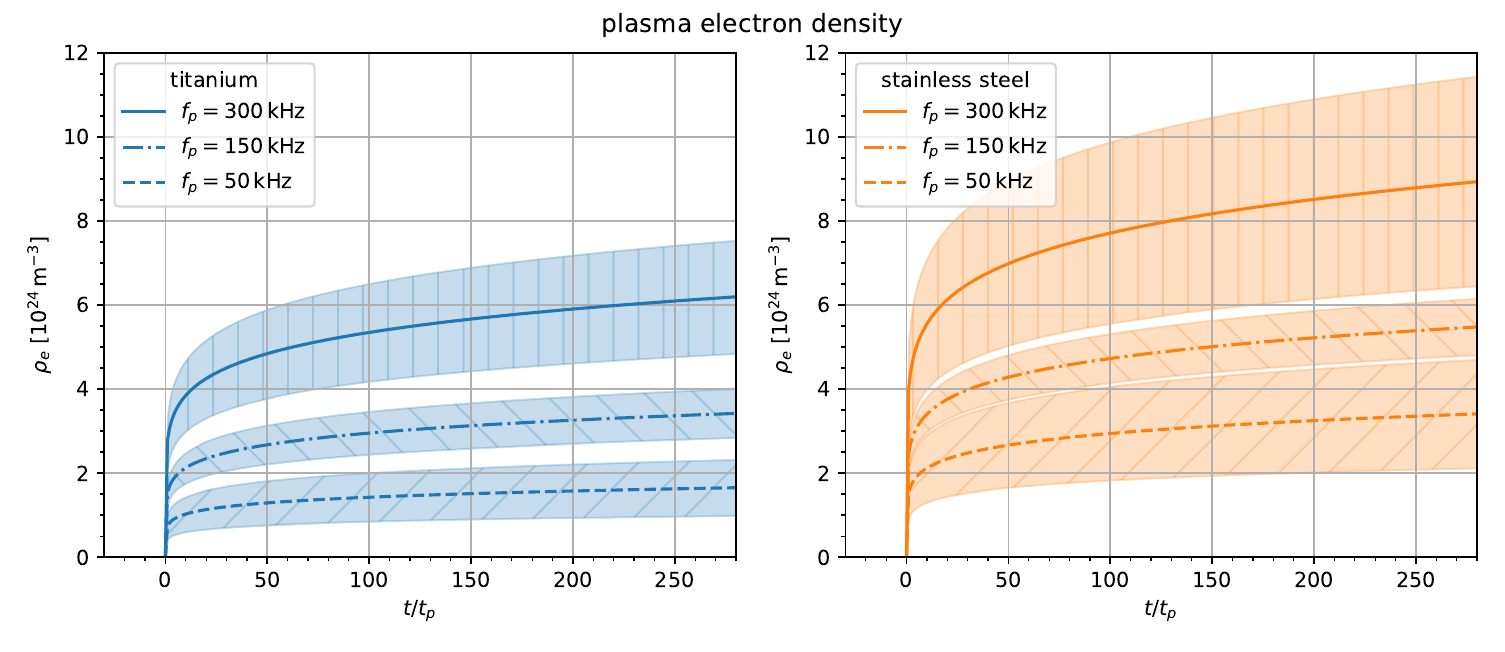}
	\caption{Plasma electronic density calculated from the average scaling constants $\epsilon$ and $\eta$ for the different experimental conditions for titanium (left) and stainless steel (right), with plume front expansion dimensionality equal to $d=1.5$. Shadowed intervals represent the corresponding standard deviations.}
	\label{fig:electronic-density}
\end{figure*}

The average electronic density $\rho_e(t)$ has been calculated as a function of time using equation \eqref{eq:electronic-density}, taking the plume front and SMI mean scaling constants $\epsilon$ and $\eta$ for the different experimental conditions. The results are reported in figure \ref{fig:electronic-density}, with the curves being plotted on the common pulse number scale, and considering the dimensionality fixed to $d=1.5$. The behavior found for $\rho_e(t)$ shows that the electronic density increases rapidly after the ablation start, and it tends to saturate between \SI{e24}{\per\meter\cubed} and \SI{e25}{\per\meter\cubed} after few hundreds of laser pulses. Higher electronic densities are observed for higher pulse frequencies, with a slightly faster increase observed for stainless steel compared to titanium.

\section{Discussion}
The experimental data presented in figure \ref{fig:experimental-exp-fit} suggest that the scaling exponents do not depend on the considered experimental conditions. This in accordance with the proposed model, where $\alpha$ and $\gamma$ depend only on the expansion dimensionality $d$. Moreover, these exponents are linked by equation \eqref{eq:exponents}, which agrees with the experimental results as it was highlighted in figure \ref{fig:exponents-scatter}. This provides a proof for the model validity in terms of only dynamical considerations, independently on possible absolute calibration errors or offsets.

The sign of $\eta<0$ confirms that the interferometric beam is mainly interacting with the plasma free electrons. In fact, a negative optical path difference~$\Delta p(t)$ can indicate a reduction in either the geometrical length or the refractive index. However, the former case can be excluded since the ablation processing is essentially superficial, and since $\Delta p(t)$ returns close to $0$ after few \si{ms}. On the other hand, a negative refraction index variation can be the effect of a gas pressure drop, as it happens behind the shock wave front, or of the presence of free electrons. The latter is typically dominant in the case of a plasma, with the refraction index of an electron gas being smaller than the vacuum value of $1$.

A clear difference between materials is present in the results for the scaling constants $\epsilon$ and $\eta$ reported in figure \ref{fig:experimental-amp-fit}. First of all, $\epsilon$ is higher for titanium. This means that the corresponding plume expansion is faster: according to equation \eqref{eq:front-radius} this may be explicable in terms of a higher shock wave energy $E_0$, as a consequence of a larger laser pulse energy transfer. On the contrary, $\eta$ is lower for titanium: this can reflect the difference observed for $\epsilon$, as it can be derived from equations \eqref{eq:eta-smi} and \eqref{eq:zeta-smi}. Indeed, the slower plume expansion observed for stainless steel as smaller $\epsilon$ values can lead to a higher plume density in a reduced volume, hence to a higher optical path difference. However, other aspects which have not been considered in the current discussion may also influence the scaling constant results, such as the atomic ionizability or other physical properties of the materials.

The presented model links the measurement from an axial interferometric probe with the average electronic density of the expanding plume. This is particularly interesting for the potential application of the SMI technique in laser technologies involving plasma formation. As a matter of fact, equation \eqref{eq:electronic-density} allows to calculate the mean electronic density $\rho_e(t)$ as a function of time from the optical path difference $\Delta p(t)$, once the plume front scaling parameters $\alpha$ and $\epsilon$ are known. The behavior of $\rho_e(t)$ calculated from experimental results and reported in figure \ref{fig:electronic-density} is consistent with expectations. In fact, the electronic density increases with time as a consequence of accumulation during a time-distributed ionization process sustained by the ablation pulse train, with a saturation effect which resembles an equilibrium with the simultaneous plume expansion. The maximum values found for $\rho_e(t)$ lay between \SI{e24}{\per\meter\cubed} and \SI{e25}{\per\meter\cubed}. Similar values have been observed by means of different techniques in other works related to laser ablation \cite{colombo_self-mixing_2017,pangovski_holographic_2016,zeng_experimental_2005,mao_simulation_2000}.

The plume accumulation phenomenon was recently observed also by an independent theoretical study \cite{ranjbar_plume_2020}. In the numerical simulations, performed in conditions similar to the ones considered here, the authors showed that in multi-pulse ablation a single shock wave arises from the pulse train, as the consequence of merging effects for the shock waves induced by the single pulses. These observations are complementary to the results presented in the current analytical and experimental study. This confirms the interpretation for plume accumulation presented here, whose preliminary quantitative characterization was reported for the first time in our previous work \cite{donadello_probing_2018}: the physical properties of the ablation plume are essentially determined by the superimposition of the multiple laser pulses. Moreover the results for the numerical simulations support the interpretation for the independence of the plume dynamics on the process parameters. Indeed, the convergence to a single shock wave is reached rapidly after few pulses, therefore the presence of the subsequent pulses generates accumulation and particle mixing within the same rarefied volume, measured by $\eta$, but does not alter significantly the expansion dynamics, measured by $\gamma$.

Quantitative considerations regarding the results for $\rho_e(t)$ should be taken carefully. Indeed, the proposed methodology might be improved by overcoming some of the assumptions which are still present in the model. First of all, plasma lifetime should be taken into account. In fact, ionized gases undergo fast recombination at standard pressures, with the electronic density decaying on scales ranging from \SIrange{e-7}{e-5}{s} \cite{verhoff_dynamics_2012,zhang_investigation_2009,choudhury_time_2016,harilal_ambient_2006}. Conversely a long-living plasma was assumed in the calculations. This may explain the differences between simulated and experimental optical paths in figures \ref{fig:ss-run1-17-plot-simulation} and \ref{fig:ss-run1-17-plot-elab}, where the experimental $\Delta p(t)$ curve undergoes a faster decay after the laser emission end. As a matter of fact, repetitive laser pulses can strongly increase the plasma lifetime within the locally rarefied plume volume, due to self-absorption and re-heating processes as it has been observed elsewhere \cite{sangines_two-color_2011,narayanan_increasing_2007,colao_comparison_2002}. Instead, after the pulse train finishes, the standard plasma lifetime determines the faster $\Delta p(t)$ decay within the vanishing plume. Also the $\rho_e(t)$ increase with the pulse repetition rate $f_p$ could be explainable in terms of such plasma lifetime enhancement. A model revision including the effects of the actual plasma lifetime might provide a more comprehensive tool for the study of plasma dynamics in ablation plumes.

The refractive index model assumed as dominant the electronic contribution, neglecting the terms related to neutral atoms and heavy ions in equation \eqref{eq:refractive-mixture}. Though for sufficiently ionized gases these contributes are typically negligible \cite{ascoli-bartoli_wavelength_1960,breitling_shadowgraphic_1999,ostrovskaya_holographic_2008,amer_laser-ablation-induced_2009}, a second order correction might be required to accurately take into account the other plume species, which may cause a $\rho_e(t)$ underestimation since they give opposite sign contributions to $\Delta n(t)$. This, in combination with the previous considerations regarding the scaling constants $\epsilon$ and $\eta$, could explain the differences between materials, with $\rho_e(t)$ being slightly higher for stainless steel compared to titanium. A possible approach that can be proposed for future studies to distinguish the refractive index contributes from the different plume species would be the implementation of a two-wavelength self-mixing interferometer. Indeed, the usage of multiple probes at different wavelengths $\lambda_{0}$ would allow to discriminate between the plume components depending on their physical properties and their kind of dependence on $\lambda_0$ \cite{ostrovskaya_holographic_2008}.

A further aspect which should require a dedicated treatment is related to the expansion dimensionality $d$, whose average value has been measured between $1$ and $2$. Due to the axial symmetry of the problem, a mainly planar propagation can be realistically assumed. The excess related to the theoretical value of $d=1$ can be interpreted as the effect of a secondary and slower transverse expansion. Moreover it must be noted that the Sedov--Taylor theory for the shock wave propagation might require modifications in the case of multi-pulse laser ablation. In fact, such distributed process cannot be approximated as a single instantaneous explosion: the initial front propagation could be perturbed by the sequence of multiple pulses, deviating from the predicted power law. Therefore the resulting plume front scaling parameters $\epsilon$ and $\alpha$ represent effective values, whose dependence on the shock wave energy $E_0$ and dimensionality $d$ introduced in equation \eqref{eq:front-radius} can be partially influenced by other parameters related to the multi-pulse laser ablation process. This may explain the effective wave energy increase which has been observed with pulse repetition rate $f_p$ in terms of a faster energy transfer, hence of bigger $\epsilon$ values. Such mechanism may also affect the expansion dimensionality on longer timescales. Conversely, $\epsilon$ and $\alpha$ have been extrapolated from the initial front expansion interval which could be acquired using the high-speed camera, hypothesizing a constant propagation mechanism. Quantitative discussions should take into account of such side effects since, together with the explicit dependence on the pulse frequency in equation \eqref{eq:eta-smi}, the behavior observed for $\epsilon$ increasing with $f_p$ may partially explain the analogous trend of $\eta$, thus of the electron number density.

\section{Conclusion}
In the current work, a method for evaluating the electron number density in a laser-induced plasma has been presented. A time-dependent analytical model describes the optical path difference in terms of refractive index variations, induced by plume formation during multi-pulse laser ablation. The model takes into account plasma accumulation and expansion given by such distributed process, describing the temporal evolution of the resulting electronic density as a power-law scaling.

The proposed method could be effectively applied to the SMI technique. Indeed, a coaxial SMI probe beam has been used in combination with a high-power pulsed laser to study the interaction with the plume generated at different pulse frequencies for titanium and stainless steel targets. A synchronous high-speed camera was used to observe the plume expansion and to calibrate the dynamical scaling parameters. The results for the scaling exponents are in agreement with the model predictions, with an average symmetry dimensionality measured as $1.5$. The electron number density was estimated from the SMI optical path difference, showing a saturation behavior between \num{e24} and \SI{e25}{\per\meter\cubed} after few hundreds of pulses, when the plasma accumulation effect tends to balance the plume propagation.

Due to the robustness and low intrusiveness of its single-arm configuration, the results showed that SMI represents a powerful and simple tool for the dynamical study of laser-induced plumes. In fact, this method could overcome the typical measurement complexity introduced by the fast evolution of such physical systems, which are of interest in many scientific fields. Moreover, SMI can be easily integrated into industrial processes which require highly time-resolved monitoring and real-time control of gas concentration. These can include laser micromachining or pulsed laser deposition, but its application might be extended to other kinds of technologies, such as ion-beam sputtering or combustion systems, where the plasma quantification is particularly important for the process optimization.

\section*{Acknowledgments}
The authors acknowledge the BLM Group for providing the acquisition setup hardware in the frame of the Project: ``LT4.0'' granted by the Regional Law 6/99 of the Autonomous Province of Trento, Italy. The Italian Ministry of Education, University and Research is acknowledged for the support provided through the Project ``Department of Excellence LIS4.0 -- Lightweight and Smart Structures for Industry 4.0''.

\bibliography{bibliography}

\end{document}